\documentclass[prl reprint, twocolumn, floatfix, preprintnumbers, showpacs]{revtex4}
\usepackage{amsfonts}
\usepackage{amsmath}
\usepackage{amssymb}
\usepackage{graphicx}

\begin{document}
\preprint{LA-UR-13-22548 }

\title{Single impurities in a Bose-Einstein condensate can make two polaron flavors}
\author{A. A. Blinova$^{1,2}$, M. G. Boshier$^{1}$, and Eddy Timmermans$^{3}$}
\affiliation{$^{1}$P-21, Physics Division, Los Alamos National Laboratory, Los Alamos, New Mexico 87545}
\affiliation{$^{2}$Physics \& Astronomy Department, Rice University, Houston, Texas 77251}
\affiliation{$^{3}$T-4, Theory Division, Los Alamos National Laboratory, Los Alamos, New Mexico 87545}


\begin{abstract}
Polarons, self-localized composite objects formed by the interaction of a single impurity particle with a host 
medium, are a paradigm of strong interaction many-body physics.  We show that dilute gas Bose-Einstein 
condensates (BEC's) are the first medium known to self-localize
the same impurity particles both in a Landau-Pekar polaron state akin to that of 
self-localized electrons in a dielectric lattice, and in a bubble state akin
to that of electron bubbles in helium.  We also show that the BEC-impurity system is 
fully characterized by just two dimensionless coupling constants, and that it can be adiabatically steered 
from the Landau-Pekar regime to the bubble regime in a smooth crossover trajectory.  
\end{abstract}

\maketitle
The polaron, a single distinguishable particle that interacts with the self-consistent deformation of the
medium that contains it, is a paradigm of strong interaction physics in condensed matter 
\cite{Devreese1972, Bogolubov2000}, chemistry \cite{Bredas1985}, and biophysics \cite{Conwell2005}.  
Polarons self-localize when sufficiently cold and strongly coupled to the host medium.  In nature, 
large \cite{RemarkLarge} polarons appear in two flavors: particles that hardly deform
the medium, such as electrons in dielectric lattices \cite{Devreese1996}, and particles that greatly 
distort the medium, such as electron bubbles in condensed helium superfluids \cite{Hernandez1991}.  
In the first class, the particle is accompanied by a cloud of small amplitude collective excitations of the 
medium which, as first shown by Landau and Pekar \cite{Landau1933, Landau1948}, can be integrated 
out in the strong coupling limit.  In the bubble systems, which occur in fluids and dense gases,  
Kuper \cite{Kuper1961} showed that the strongly repelling particle can be described as residing in a 
self-created cavity -- the ``bubble'' -- surrounded by the fluid.  The effective mass and mobility of bubble 
and Landau-Pekar polarons exhibit quite different behaviors, and so the two polaron flavors are customarily 
treated as distinct.  However, as we show below, dilute gas Bose-Einstein condensates (BECs)  
\cite{Anderson1995, Davis1995}, unlike previously known host media, can self-localize neutral impurity atoms
\cite{Cucchietti2006, Kalas2006} in both the bubble and Landau-Pekar polaron states.  Further, the 
system phase diagram presented here shows how the strongly coupled BEC-impurity polaron evolves 
continuously between these limits as the interaction strengths and BEC density are varied.  
The BEC-impurity system can therefore be regarded as a quantum simulator of large
polarons in a boson environment \cite{RemarkFermion}, complementing recent proposals 
for simulating lattice polarons \cite{Mathey2004, Bruderer2007, Stojanovic2012, Mezzacapo2012}.
 
\textit{The system:}   We consider a neutral impurity atom of mass $m_{I}$ immersed in a homogeneous
BEC of $N$ boson particles of mass $m_{B}$ contained in a macroscopic
volume $\Omega$, giving an average density $\rho=N/\Omega$.  Bosons
at positions ${\bf r}$ and ${\bf r}'$ interact via a repulsive short-range
interaction of scattering length $a_{BB}$, described by an effective potential
$V_{BB}\left( {\bf r}-{\bf r}'\right) = \lambda_{BB} \delta \left( {\bf r}-{\bf r}'\right)$,
where $\lambda_{BB}=\left( 4 \pi \hbar^{2} / m_{B} \right) a_{BB}$ and $a_{BB}
> 0$.  The boson chemical potential $\mu_{B}=\lambda_{BB} \rho$ sets the 
time scale $\hbar/\mu_{B}$ and the coherence length $\xi = 1/\sqrt{16 \pi \rho a_{BB} }$ sets the length scale on which the BEC can 
respond to a perturbation.  The perturbation here is provided by the 
impurity interacting with the bosons via $V_{IB}\left( {\bf r}-{\bf x}\right)
= \lambda_{IB} \delta \left( {\bf r}-{\bf x}\right)$, where ${\bf x}$
represents the impurity position and the impurity-boson interaction
strength $\lambda_{IB}=2 \pi \hbar^{2} \left( m_{I}^{-1}
+ m_{B}^{-1} \right) a_{IB}$ is proportional to the impurity-boson scattering
length $a_{IB}$, taken to be Feshbach-tuned \cite{Chin2010} to a large positive value \cite{Bruderer2008}. 
We break the translational symmetry of the BEC-impurity system ground state by fixing the impurity 
center-of-mass position at ${\bf r}=0$ \cite{RemarkFixing}.  We write the BEC density in the presence 
of the impurity as $\rho_{B}\left({\bf r}\right) = \rho + \delta \rho_{B} \left({\bf r}\right)$.
Self-localization occurs when the effective potential $\lambda_{IB} \delta \rho_{B} \left( {\bf r} \right)$ can 
trap the impurity.  We classify $ \left|\delta \rho_{B}({\bf r}=0)/\rho \right| <  0.1$ as the Landau-Pekar 
regime \cite{Cucchietti2006}, and $\left| \delta \rho_{B}({\bf r}=0)/\rho \right| > 0.9$ as the bubble regime.   
We will show that in these regimes the impurity observables exhibit the scaling behaviors expected 
from the simplified Landau-Pekar and bubble descriptions respectively.

\begin{figure}
\includegraphics[width=3.25in, trim=0.3in 0.3in 0.3in 0]{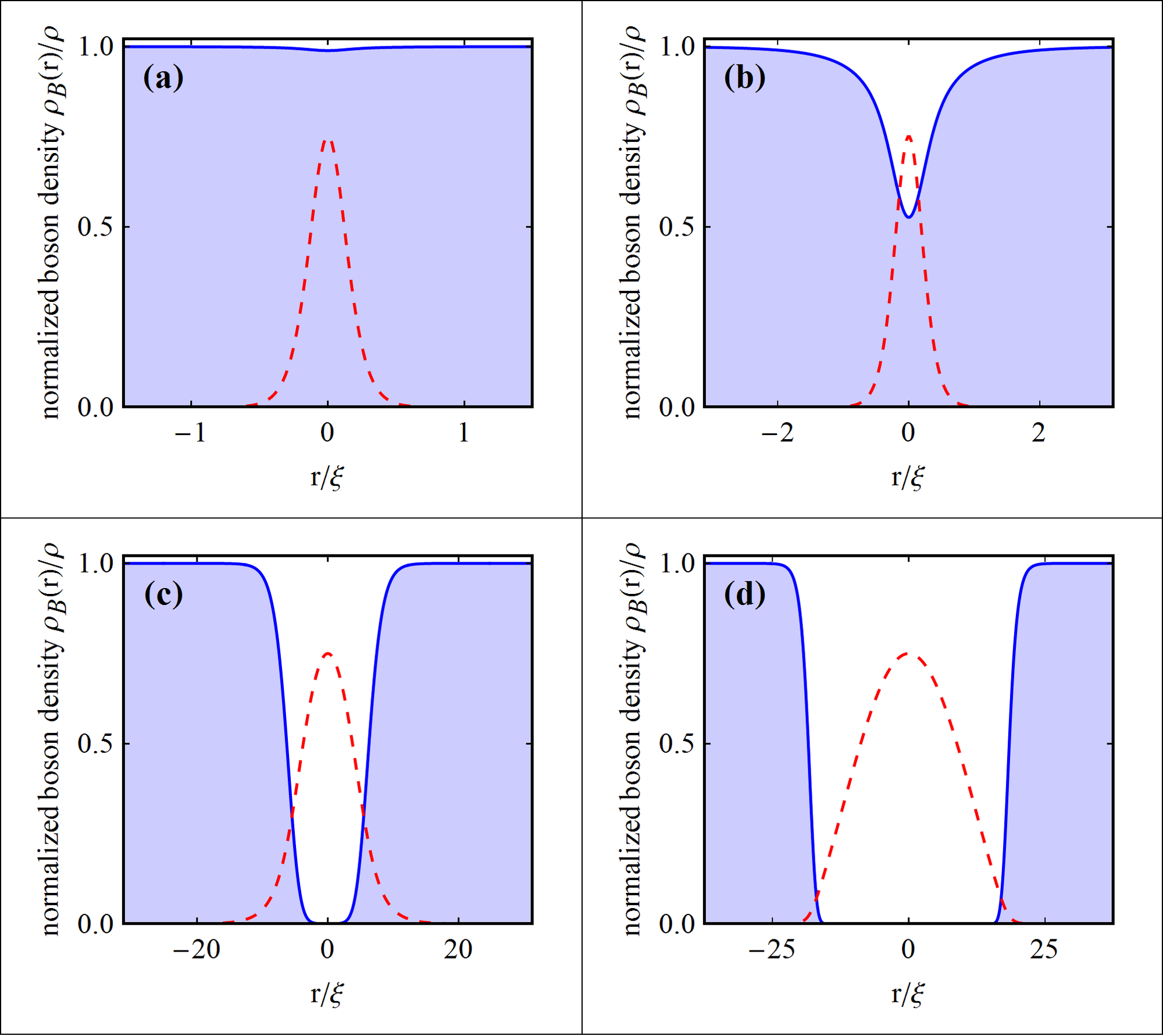}
\caption{\label{bosonProfile} Numerical results for the normalized boson density (blue shading) and 
un-normalized impurity wavefunction (red dashed line).  Parameter values [see Eqs.\,(\ref{e:beta}) 
and (\ref{e:alpha})]: (a) $\beta = 25$ and $\alpha = 10^{-9 }$, (b) $\beta = 25$ and $\alpha = 10^{-5 }$, 
(c) $\beta = 25$ and $\alpha = 10^{1 }$, (d) $\beta = 5 \times 10^{4}$ and $\alpha = 10^{3 }$. }
\end{figure}

\textit{Landau-Pekar polaron: }The impurity-boson repulsion can be simultaneously strong enough to self-trap 
the impurity, and weak enough to hardly change the BEC density profile [Fig.\,\ref{bosonProfile}(a)].  
In this regime, the Bogoliubov expansion and transformation that describes the BEC fluctuations \cite{Bogoliubov1947}
can be carried out around the homogeneous BEC \cite{Santamore2011}.  The interaction
of the dilute BEC with an impurity of density $\rho_{I}
\left({\bf r}\right) = \left( 2 \pi\right)^{-3} \int d^{3} {\bf k} 
e^{i {\bf k} \cdot {\bf r}} \rho_{I,{\bf k}}$ then gives a Fr{\"o}hlich
hamiltonian \cite{Froehlich1954}, familiar from electron-phonon 
interactions.  Representing the quasi-particle annihilation/creation
operators of momentum ${\bf k}$ and energy $\hbar \omega_{k}=
\hbar k \sqrt{(1 + \xi^{2} k^{2})(\mu_{B}/m_{B})}$
by $b_{\bf k},b_{\bf k}^{\dagger}$, the impurity-boson interaction
is described by
\begin{align}
H_{IB} &= \frac{\lambda_{IB}}{\Omega} \sum_{\bf k} \rho_{I,-{\bf k}}\; \rho_{B,{\bf k}}\nonumber{}\\
           &\approx \lambda_{IB} \rho + \frac{M}{\sqrt{\Omega}} \sum_{\bf k} \nu_{k} \rho_{I,-{\bf k}} \left( b^{\dagger}_{\bf k} + b_{\bf k} \right) \;, 
\label{e:bog}
\end{align}
where $\rho_{B,{\bf k}}$ is the operator associated with the boson density.  In Eq.\,({\ref{e:bog}), 
the impurity-phonon interaction matrix element $M=\lambda_{IB} \sqrt{\rho}$ , and 
$\nu_{k}=\left( \xi^{2} k^{2} / \left[1 + \xi^{2} k^{2} \right] \right)^{1/4}$ is a 
structure factor arising from the Bogoliubov transformation.  By showing that the 
Fr{\"o}hlich coupling $H_{IB}$ amounts to a displacement of the ${\bf k}$-mode oscillator coordinate
$\phi_{\bf k} = \left( b_{\bf k}^{\dagger} + b_{-{\bf k}} \right)/\sqrt{2}$,
Landau and Pekar \cite{Landau1948} integrated out the phonon modes.
The resulting energy reduction takes the form of a self-interaction \cite{Santamore2011, Casteels2011},
$\Delta E = -M^{2} \left(2\pi \right)^{-3} \int d^{3} {\bf k} \rho_{I,-{\bf k}}
\rho_{I,{\bf k}} \left( \nu_{k}^{2}/\hbar \omega_{k} \right)$, which in the strongly coupled regime 
overcomes the kinetic energy cost of localizing the impurity.  Here, the self-interaction potential
is an attractive Yukawa potential of range $\xi$.  It can cause self-localization when $\xi$ exceeds the 
impurity extent, which is comparable to \cite{Cucchietti2006}
\begin{equation}
R_{o} = \left[4\pi \rho  a_{IB}^{2} \left( 1 + m_{I}/m_{B} \right) \left( 1 +\ m_{B}/m_{I} \right)\right]^{-1} \; .
\label{e:ro}
\end{equation}
Specifically, when the ratio of the boson healing length $\xi$ to  the self-localization length $R_{o}$
\begin{equation}
\beta=\frac{\xi}{R_{o}} = \sqrt{\pi \rho \frac{a_{IB}^{4}}{a_{BB}}} 
\left( 1 + \frac{m_{I}}{m_{B}} \right) \left( 1 + \frac{m_{B}}{m_{I}} \right) \geq 5, 
\label{e:beta}
\end{equation}
the above description predicts self-localization \cite{Cucchietti2006}.  Note that the impurity
extent $R_{o}=\xi/\beta$ can be significantly smaller than $\xi$, making this polaron 
a sub-coherence length structure.  The binding energy, proportional to
\begin{equation}
E_{o} = \frac{ \hbar^{2}}{ 2 m_{I} R_{o}^{2} } = 2 \left( \frac{m_{B}}{m_{I}} \right) 
\beta^{2} \mu_{B}, 
\end{equation}
then significantly exceeds $\mu_{B}$.  When scaled by $E_{o}$ and $R_{o}$,
impurity observables in the Landau-Pekar regime depend only on the dimensionless coupling 
strength $\beta$ \cite{Cucchietti2006}.

\textit{Bubble polaron:}  When $a_{IB}$ grows sufficiently to expel the BEC from the impurity's 
vicinity [Fig.\,\ref{bosonProfile}(c) and (d)], a Bogoliubov procedure should expand around the deformed BEC.
However, a bubble description such as Kuper's model of electron bubbles in helium is a simpler starting point \cite{Kuper1961}.
With complete BEC and impurity separation [Fig.\,\ref{bosonProfile}(d)] the impurity, trapped in a 
self-created spherical cavity of radius $R_{c}$ and volume $V_{c}$, has wavefunction 
$\chi \left( r \right) = \sqrt{\pi R_{c}}^{-1}\sin\left( \pi r/R_{c} \right)/r$.  Neglecting surface tension, 
the system energy difference $E_{c}$ with and without impurity is the impurity kinetic energy
$\pi^{2} \hbar^{2}/\left(m_{I} R_{c}^{2} \right)$ plus the 
energy cost of making the cavity, $P V_{c}$, where  $P= \lambda_{BB} \rho^{2}/2$ is
the BEC pressure.  Hence
\begin{equation}
E_{c}(R_{c}) = \frac{\pi^{2} \hbar^{2}}{m_{I} R_{c}^{2}} 
+ \frac{8 \pi^{2}}{3} \frac{\hbar^{2} a_{BB}}{m_{B}} \rho^{2} R_{c}^{3} \; .
\end{equation}
The minimization $\partial E_{c}/\partial R_{c} = 0$ yields the expected cavity radius
$R_{c}= \left[ 4\left(m_{I}/m_{B}\right) \rho^{2} a_{BB} \right]^{-1/5}$, and the impurity energy
 $E_{c} = (5/3) \left( \pi^{2} \hbar^{2} / \left[ m_{I} R_{c}^{2} \right]\right) 
= \left( 5 \pi/2^{11/5}\right) \left(m_{B}/m_{I}\right)^{3/5} \mu_{B}/ \left[ \sqrt{\rho a_{BB}^{3}} \right]^{2/5}$.  

\textit{BEC permeability:} The radical change in BEC-impurity overlap seen in Fig.\,\ref{bosonProfile}(a) - (d) 
is due to a BEC ``stiffness'' arising because it costs kinetic and interaction energy to move 
$\Delta N_{B} = \left| \int d^{3} {\bf r} \left[ \rho_{B}\left({\bf r}\right) - \rho \right] \right|$
bosons away from the impurity. 
Estimating the interaction energy cost as $E_{x} = \Delta N_{B} \mu_{B}$
and using the predicted displaced boson number from \cite{Massignan2005},
$\Delta N_{B} = \left| \lambda_{IB}/\lambda_{BB} \right|$ (valid in the Landau-Pekar
and crossover regimes but not in the bubble regime), gives 
$E_{x} = \left| \lambda_{IB}/ \lambda_{BB} \right| \mu_{B}$.  The ratio 
\begin{equation}
\sigma = \frac{E_{x}}{E_{o}}
= \left[ 4 \pi \rho a_{IB}^{3} \left( 1 + \frac{m_{I}}{m_{B}}\right)^{3}\left(\frac{m_{B}}{m_{I}}\right)^{2}\right]^{-1}
\label{e:sigma}
\end{equation}
then quantifies the relative importance of the displacement energy cost to the overlap energy 
gain of self-localization.  We refer to $\sigma$ as the BEC-impurity ``permeability''.  
In the Landau-Pekar regime, a direct analytical evaluation yields 
$ \left| \delta \rho_{B} \left( {\bf r}=0 \right)/\rho \right| = \left( 4 \sqrt{2}/3 \sqrt{\pi} \right) \sigma^{-1} = 1.064\,\sigma^{-1}$.
Thus, $\sigma \gg 1$ implies Landau-Pekar conditions where the repulsion is insufficient to 
overcome the BEC stiffness and displace the bosons noticeably [Fig.\,\ref{bosonProfile}(a)].  
A gradual increase in $a_{IB}$ then expels the bosons significantly when 
$\sigma \sim 1$ [Fig.\,\ref{bosonProfile}(b)], and enters the large-depletion bubble limit 
[Fig.\,\ref{bosonProfile}(c) and (d)] when $\sigma \ll 1$.

\textit{General case:}  A more general ground state treatment, encompassing the Landau-Pekar
and bubble regimes as limits, is based on the strong coupling approximation of a many-body product state.  
Minimizing the energy while requiring the respective boson and impurity wavefunctions 
$ \psi \left({\bf r}\right)$ and $\chi \left({\bf r}\right)$ to be normalized gives two
coupled Gross-Pitaevskii equations
\begin{eqnarray}
\label{e:psichi}
\mu_{B} \psi \left({\bf r}\right) &=&
- \frac{\hbar^{2} \nabla^{2}}{2m_{B}} \psi\left({\bf r}\right) + \lambda_{BB} \left| \psi\left({\bf r}\right)\right|^{2} \psi\left({\bf r}\right)\nonumber\\
&&+ \lambda_{IB} \left| \chi\left({\bf r}\right)\right|^{2} \psi \left({\bf r} \right) 
\\
\epsilon_{I} \chi\left({\bf r}\right) &=&
- \frac{\hbar^{2} \nabla^{2}}{2m_{I}} \chi \left({\bf r}\right) + \lambda_{IB} \left| \psi\left({\bf r}\right)\right|^{2} \chi \left({\bf r} \right),\nonumber 
\end{eqnarray}
where  $\lim_{r\rightarrow \infty} \psi(r)= \sqrt{\rho}$,  and $\mu_{B}$ and $\epsilon_{I}$ 
represent the Lagrange multipliers ensuring normalization, 
$\int d^{3} {\bf r} \; \left| \psi \left( {\bf r} \right) \right|^{2} = N$, 
$\int d^{3} {\bf r} \; \left| \chi \left( {\bf r} \right) \right|^{2} = 1$.

The BEC-impurity system has five physical parameters --
$m_{B}, m_{I}, \rho, a_{BB}$ and $a_{IB}$ -- but we find that the proper dimensional
scaling of energies, density, and distances reveals a minimal dependence
on just two coupling constants.  The first is the length scale ratio
$\beta = \xi/R_{o}$.   The second is the mass-scaled boson gas parameter
\begin{equation}
\alpha =\left( \frac{m_{B}}{m_{I}} \right) \sqrt{ \rho a_{BB}^{3}} \; .
\label{e:alpha}
\end{equation}
All properly scaled observables can be cast in terms of $\alpha$ and $\beta$.  For example, 
the permeability parameter $\sigma$ takes the form 
$\sigma\left(\alpha,\beta\right) = 1/(4 \pi^{1/4} \sqrt{\alpha \beta^{3}})$, 
and in the bubble limit the energy and cavity radius are given by 
$E_{c}=E_{o} \left( 5 \pi/24 \right) \beta^{2} \left(4 /\alpha \right)^{2/5}$ 
and
$R_{c} = R_{o} 4 \pi^{1/2} \beta \alpha^{1/5}$, 
respectively. To prove the minimal dependence, we substitute (real-valued) 
scaled dimensionless boson ($p$) and impurity ($g$) wavefunctions:
\begin{eqnarray}
\psi \left( {\bf r} \right) &=& \sqrt{\rho} \; p \left({\bf x} = {\bf r}/\xi \right)
\nonumber \\
\chi \left( {\bf r} \right) &=& R_{o}^{-3/2}  g\left({\bf y} = {\bf r}/R_{o}\right) \; ,
\label{e:gp}
\end{eqnarray}
into the coupled equations (\ref{e:psichi}), scaling the energies and lengths to 
obtain the dimensionless form 
\begin{multline}
\left( \nabla_{x}^{2} + \frac{1 - p^{2}\left({\bf x}\right)}{2} \right) p\left({\bf x}\right)
= - \frac{4 \pi \beta^{2}}{\sigma\left(\alpha,\beta\right)} g^{2}\left({\bf y}={\bf x}\beta\right) p\left({\bf x}\right)
 \\
\left( \nabla_{y}^{2} + e_{I} \right) g\left({\bf y}\right)
= - \sigma\left(\alpha,\beta\right) p^{2}\left({\bf x}={\bf y}/\beta\right) g\left({\bf y}\right),
\label{e:scaled}
\end{multline}
where $e_{I}=\epsilon_{I}/E_{o}$, and the $g, p$ solutions have to satisfy the normalization
$\int d^{3} {\bf y} g^{2} \left({\bf y}\right)=1$ 
and boundary condition
$\lim_{x\rightarrow\infty} p\left({\bf x}\right)=1$.  
Since the scaled coupled equations depend only on $\alpha$ and 
$\beta$ \cite{RemarkScaled} the BEC-impurity phase diagram is two-dimensional (2D).

Numerical solutions of the coupled equations (\ref{e:psichi}) provide a rigorous test of the 
Landau-Pekar and bubble limiting behaviors by determining the $\alpha$- and $\beta$-dependences 
of the relevant  observables.  For example, in Fig.\,\ref{ImpuritySize} we
plot the root mean-square impurity extent 
$R_{rms}=\sqrt{ \int d^{3} {\bf r}\,r^{2} \rho_{I}({\bf r})}$.   
Each curve shows $R_{rms}/R_{o}$ as a function of $\alpha$ for a specific $\beta$-value 
on a log-log plot, with arrows indicating $\alpha$ values corresponding to unit permeability $\sigma=1$.  
On the left, the $\alpha$-independence of the Landau-Pekar impurity properties is confirmed
by the zero slope of the curves, which also show the expected convergence to 
$R_{rms} \approx 4.6\,R_{0}$  for $\beta \ge 20$ \cite{Cucchietti2006}.  On the right, the straight 
lines of slope $1/5$ confirm the expected bubble scaling 
$R_{rms}/R_{o}\propto \beta \alpha^{1/5}$.  
The smooth change between these limits indicates a crossover, rather than a transition, 
between Landau-Pekar and bubble regimes.  
\begin{figure}
\includegraphics[width=3.25in, trim=0 0.25in 0 0.5in]{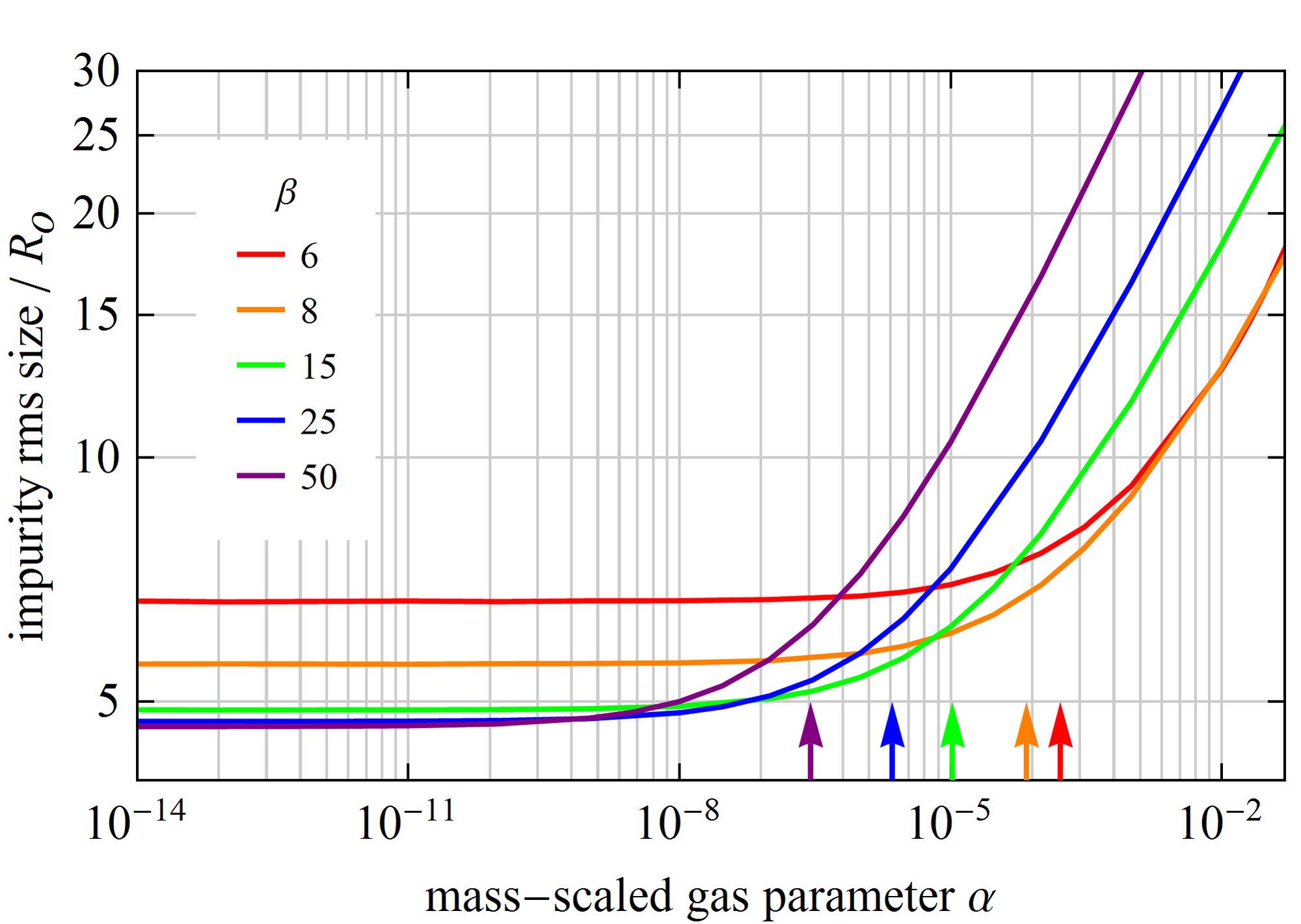}
\caption{\label{ImpuritySize} Dependence on gas parameter $\alpha$ of the r.m.s. width 
of the impurity density (in units of scaled length $R_0$), for several values of impurity-BEC 
interaction parameter $\beta$.    For each $\beta$ arrows indicate the values of $\alpha$ 
for which permeability $\sigma =1$.}
\end{figure}
\begin{figure}
\includegraphics[width=3.25in, trim=0 0.25in 0 0.5in]{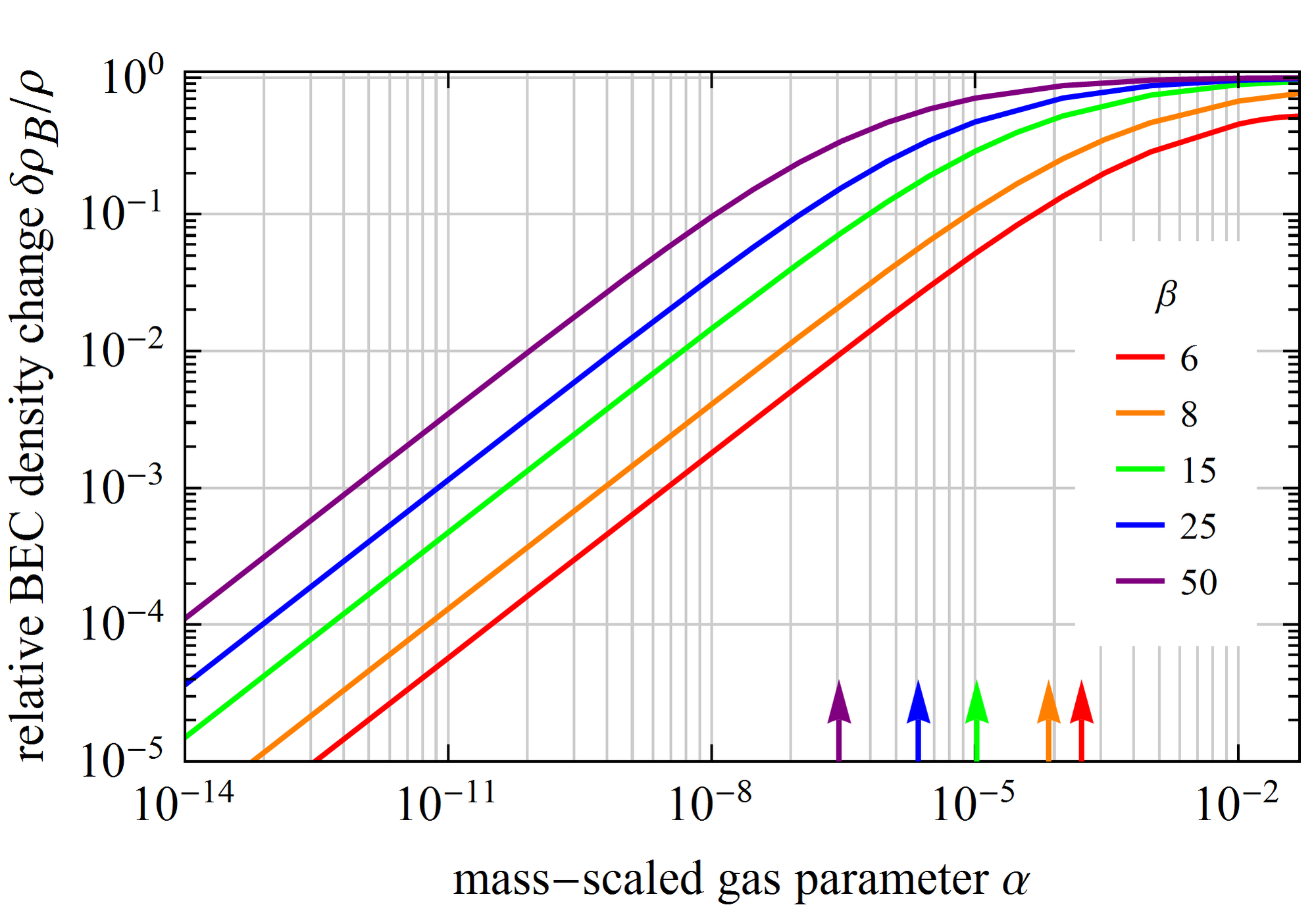}
\caption{\label{BosonDepth} Dependence on gas parameter $\alpha$ of the relative 
decrease in BEC density at the impurity location, for several values of impurity-BEC 
interaction parameter $\beta$.  For each $\beta$ arrows indicate the values of $\alpha$ 
for which permeability $\sigma =1$.}
\end{figure}

Likewise, the relative BEC density change at the impurity position, 
$\left| \delta \rho_{B}\left( {\bf r} = 0\right) /\rho \right|$, 
plotted in Fig.\,\ref{BosonDepth} shows the 
$\sigma^{-1}\propto \alpha^{1/2}$ 
scaling of the Landau-Pekar polaron on the left side.  Before crossing the line of unit relative 
density response, the curves level off and approach the bubble limit of maximal BEC depletion asymptotically.
As in Fig.\,\ref{ImpuritySize}, the arrows on Fig.\,\ref{BosonDepth} indicating the $\sigma=1$ positions 
illustrate that the crossovers occur near unit permeability.  Fig.\,\ref{PhaseDiagram} is the 2D 
phase diagram for the BEC-impurity system, produced by plotting 
$\left| \delta \rho_{B}({\bf r}=0)/\rho\right|$ 
in the ($\alpha, \beta$)-plane, colored to show the polaron regimes and the region where the impurity-BEC interaction is not strong enough to cause self-localization.  The red $\sigma=1$ line lies, as expected, on top of the crossover region. 

\textit{Experimental realization:}  The remarkable ability of a BEC to self-localize impurities both 
in Landau-Pekar and in bubble states could be strikingly illustrated by an experiment
that adiabatically Feshbach-tunes the same BEC-impurity system from one
limit to the other.  For typical densities 
($10^{11}\,\textrm{cm}^{-3} < \rho < 10^{14}\,\textrm{cm}^{-3}$) 
and realistic ranges for $a_{BB}$ and $m_{B}/m_{I}$, we find that cold atom $\alpha$ values may range from 
$10^{-7}$ to $10^{-1}$.  
Increasing $a_{IB}$ by Feshbach tuning could achieve self-localization at 
$\beta \sim 5$ (see Fig.\,\ref{PhaseDiagram}).  
Using Eq.\,(\ref{e:beta}) and scaling scattering lengths and densities by the typical values of 
1\,nm and $\overline{\rho}=10^{13}\,{\textrm{cm}}^{-3}$, this corresponds to 
$a_{IB}=a_{IB}^{s.l.}$, where
\begin{equation}
a_{IB}^{s.l.}=\frac{168\, \textrm{nm}}{\sqrt{ \left( 1 + \frac{m_{B}}{m_{I}} \right) \left( 1 + \frac{m_{I}}{m_{B}} \right)}}
\left( \frac{ a_{BB}/\textrm{nm}}{\rho/\overline{\rho}} \right)^{1/4} .
\label{e:selfloc}
\end{equation}
\begin{figure}
\includegraphics[width=3.25in, trim= 0 0 0 0.5in]{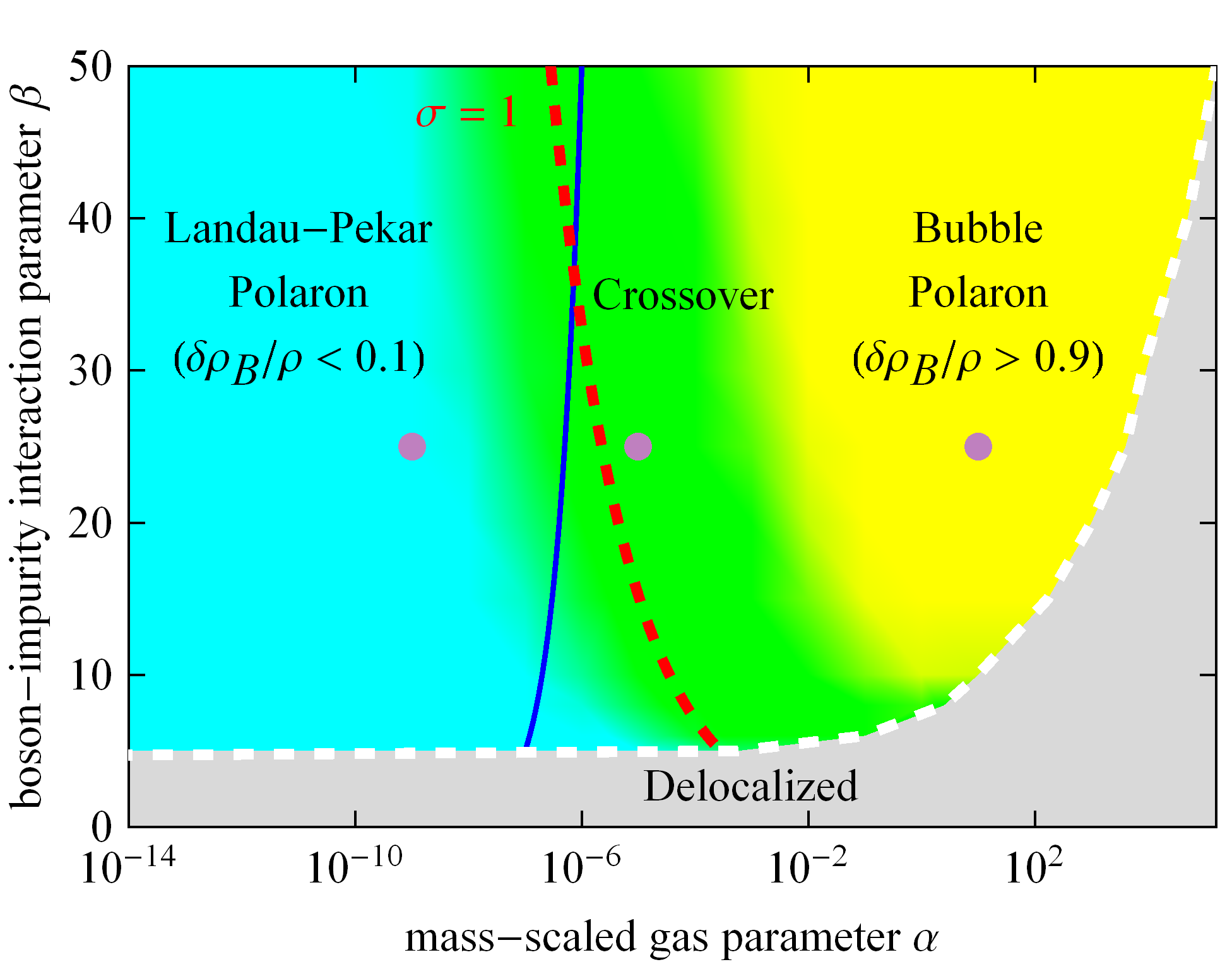}
\caption{\label{PhaseDiagram} Phase diagram of the BEC-impurity system obtained as a plot of 
$\left| \delta\rho_{B}({\bf r}=0)/\rho \right|$.  The dots in the Landau-Pekar, crossover, 
and bubble regions correspond to the BEC density profiles in 
Fig.\,\ref{bosonProfile}(a), (b), and (c), respectively.  The blue line is a trajectory that tunes the system 
continuously from Landau-Pekar limit to crossover region to bubble limit (off the top of the plot).}
\end{figure}
This self-localization near $\beta \sim 5$ results in a Landau-Pekar polaron if 
$\sigma=\left[ 4 \pi^{1/4} \sqrt{125 \alpha}\right]^{-1}\gg 1$, requiring
\begin{equation}
a_{BB} \ll 2.00 \,\textrm{nm}\left(\rho/\overline{\rho}\right)^{-1/3} \left(m_{B}/m_{I}\right)^{-2/3}\; .
\end{equation}
A further increase in $a_{IB}$ and/or $\rho$ can lower the permeability and effect a 
cross over to the bubble regime when $a_{IB} \sim a_{IB}^{cross}$, where from Eq. (\ref{e:sigma})
\begin{equation}
a^{cross}_{IB} = 200\,\textrm{nm} \frac{\left(\rho/\overline{\rho}\right)^{-1/3} \left( m_{B}/m_{I}\right)^{-2/3}}{
\left(1 + m_{I}/m_{B}\right)} \; .
\end{equation}
Throughout, the permeability parameter varies as
\begin{equation}
\sigma = \frac{1.68}{\left(a_{IB}/\overline{a}_{IB}^{s.l.}\right)^{3}} 
\frac{1}{\left(\rho/\overline{\rho}\right)^{1/4}} \frac{1}{\left(a_{BB}/1\,\textrm{nm}\right)^{3/4}} \left(\frac{m_{B}}{m_{I}} \right)^{-\frac{1}{2}},
\end{equation}
where $\overline{a}_{IB}^{s.l.}$ denotes the scattering length of Eq.(\ref{e:selfloc}) at standard density, $\rho=\overline{\rho}$.
The adiabatic $\left( a_{IB},\rho\right)$ variation traces out a trajectory on the phase diagram 
(e.g. blue line on Fig.\,\ref{PhaseDiagram}) that starts in the low $\alpha$, cyan-colored, Landau-Pekar region.  
An increase in BEC density (corresponding to an exponential increase in $\left( \ln \alpha, \beta \right)$-
space) and a Feshbach increase in $a_{IB}$ (corresponding to a vertical upward motion in
the same space) can then, eventually, steer the BEC-impurity across the green-colored crossover regime into
the yellow-colored bubble region.

We now consider two potential issues facing such an experiment.  The first is the lifetime of the impurity 
against three-body recombination.  An increase in $a_{IB}$ is generally (but not always -- see below) 
accompanied by a decrease in lifetime \cite{Inouye1998}.  Estimating
the three-body limited impurity lifetime $\tau_{I}$ as in \cite{Fedichev1996}, we expect
$\tau_{I}^{est} \sim \left(\sqrt{3}/3.9\right) \left[ \sqrt{1 + 2 \left(m_{B}/m_{I}\right)}\left( \hbar/m_{B}\right) a_{IB}^{4} \rho^{2} \right]^{-1}$.  
As the time scale of the BEC response (and the slowest time scale in the system), we expect 
$\tau_{B}=\hbar/\mu_{B}$
to set the scale of the self-localization dynamics.  The estimated impurity lifetime with
full overlap can then significantly exceed $\tau_{B}$ as long as $\beta$ is not too large,
\begin{equation}
\tau_{I}^{est} = \tau_{B} \frac{4\pi^{2}}{\beta^{2}} \frac{\sqrt{3}}{3.9} 
\frac{ \left( 1 + m_{I}/m_{B} \right)^{2} \left( 1 + m_{B}/m_{I} \right)^{2} }
{\sqrt{ 1 + 2 m_{B}/m_{I}}} \; .
\end{equation}
A more careful study of the three-body loss \cite{Esry1999} found that 
$\tau_{I}^{est}$ should be divided by an oscillating Stuckelberg factor 
related to three-body Efimov physics \cite{Wang2012}.  Near the nodes of the Stuckelberg
factor the lifetime greatly increases.   Further, the near complete separation of the impurity 
and the BEC in the bubble limit implies a significant increase in impurity lifetime for sufficiently large 
$a_{IB}$ because the three-body recombination loss rate is proportional to the overlap 
$\int d^{3} {\bf r} \; \rho_{B}^{2} \left({\bf r} \right) \rho_{I}\left({\bf r}\right)$.  
We note that in condensed helium, the increased lifetime of positronium \cite{Paul1957, Wackerle1957} 
has been used as a signal of self-localization \cite{Ferrell1958}.

The second challenge faced by an experimental realization of the bubble polaron may be the buoyancy force 
${\bf F} = - \nabla E_{c}\left(\rho\right) = - \left( 4 /5 \right) E_{c} \left( \nabla \rho\right)/\rho$ 
attempting to expel the impurity from the high density region in an inhomogeneous BEC.  A two-color trap 
or a species specific potential \cite{Catani2009, Lamporesi2010} may be necessary to keep the impurity 
bubble in place. Alternatively, a homogeneous condensate might be created in a flat-bottomed 
Painted Potential \cite{Henderson2009}.  We note that while the bubble polaron always seeks low BEC density, 
the Landau-Pekar polaron can be high density seeking.

\textit{Conclusion:}  We have shown that a distinguishable neutral atom embedded in a 
dilute BEC can, if the BEC-impurity repulsion is strong enough, self-localize in a bubble polaron
state in which the impurity is impermeable to the BEC host medium.  This state
is analogous to that of an electron bubble in condensed helium.  Remarkably, and
uniquely, the same BEC medium can also self-localize impurities in Landau-Pekar
polarons if the BEC density and interaction strengths have appropriate values.  We characterized 
the overlap of the self-localized impurity with the host fluid by a permeability parameter $\sigma$ 
that ranges from $\sigma \gg 1$ in the Landau-Pekar limit to $\sigma \ll 1$ in the bubble limit.  
We have shown that the BEC-impurity system is fully characterized by just two dimensionless 
coupling constants.  In the corresponding phase diagram, the bubble and Landau-Pekar states 
correspond to broad regions that are separated by a smooth crossover region near $\sigma \sim 1$.
Finally, we pointed out that a single impurity-BEC system can be adiabatically steered from one 
limit to the other. 
 
E.T. would like to thank the Aspen Center for Physics for a visit, during which part of this work was conceived.  This work was partially funded by the Los Alamos National Laboratory LDRD Program.


\begin{thebibliography}{99}
\bibitem{RemarkFixing} The true ground state preserves the translational symmetry, but the symmetry breaking energy cost is negligible in the strong-coupling limit.
\bibitem{RemarkScaled} $\beta$ and $\sigma$ can also be used as coupling parameters, but we prefer to use $\alpha$ because tuning $a_{IB}$ leaves $\alpha$ unchanged.
\end{thebibliography}
\end{document}